\documentclass[prd,onecolumn,amsfonts,amssymb,superscriptaddress,notitlepage]{revtex4-1}

\usepackage{color}
\usepackage{amsmath}

\pdfoutput=1

\begin{document}


\title{On the existence of stationary Ricci solitons}

\author{Pau Figueras}
\affiliation{School of Mathematical Sciences, Queen Mary University of London, Mile End Road, London E1 4NS, UK} 
\author{Toby Wiseman}
\affiliation{Theoretical Physics Group, Blackett Laboratory, Imperial College, London SW7 2AZ, UK }


\date{October 2016}

\begin{abstract}
Previously the DeTurck `trick' has been used to render the stationary Einstein's equation a well posed elliptic system that may be solved numerically by geometric flow or directly. 
Whilst in the static case for pure gravity with zero or negative cosmological constant there is a simple proof that solving the modified ``harmonic'' Einstein's equation leads to a solution of the original Einstein system -- i.e. not a Ricci soliton -- in the stationary case this argument no longer works.
Here we provide a new argument that extends the static result to the case of stationary spacetimes that possess a ``$t$-$\phi$" reflection symmetry. 
Defining a ``soliton charge'' from the asymptotic behaviour of the solution, we show that 
this quantity is always non-positive.
Provided asymptotic conditions are chosen such that this charge vanishes, then stationary solitons cannot exist.

\end{abstract}

\maketitle

%
\section{Introduction}
%

Solutions to Einstein's equation of general relativity are notoriously difficult to find but they have played a prominent role in the understanding of the theory and its physical implications. Amongst all solutions, black holes are especially important since they capture some of the key features of Einstein's theory of gravity, namely the fact that spacetime has a non-trivial causal structure and that singularities are present in their interior where the theory breaks down. 

In the event that general relativity is indeed the correct classical description of gravity, then Kerr's celebrated solution \cite{Kerr:1963ud} describes the steady state of \textit{all} astrophysical black holes in the Universe, and the recent detection of gravitational waves has provided direct experimental evidence for the first time supporting this \cite{Abbott:2016blz,Abbott:2016nmj}. Kerr's solution is very remarkable in many regards. In particular, it depends on only two physical parameters, mass and angular momentum, and is highly symmetric. More precisely, the Kerr spacetime possesses two commuting Killing vector fields\footnote{The Kerr spacetime is also algebraically special, namely Petrov type D.} which underlies why its explicit form is known. In four dimensions, Einstein's vacuum equation for spacetimes possessing two commuting Killing vector fields (the maximal number possible for asymptotically flat solutions) is integrable and all solutions within this symmetry class can be found analytically by an entirely algebraic procedure.

In recent years, motivated by string theory,  there has been a lot of interest in finding black hole solutions in Einstein's gravity and beyond in dimensions higher than four and/or with the presence of a cosmological constant and other matter fields. However, beyond the astrophysical setting, the integrability of Einstein's equation is generically lost and new techniques are needed in order to find black hole solutions analytically \footnote{In general, for vacuum spacetimes for $D-2$ commuting Killing vector fields the Einstein's equation is integrable, where $D$ is the total number of spacetime dimensions. Note that only in $D=4,\,5$ these spacetimes can be asymptotically flat.}. In certain supergravity theories one can make progress by imposing supersymmetry, but this only allows one to probe a small subset of solutions. 

On the other hand, progress has been made in solving Einstein's equation using numerical methods. Ref. \cite{Wiseman:2001xt} was the first to propose a method to solve Einstein's equation for co-homogeneity two spacetimes which do not possess the maximal number of commuting Killing vectors. Making a particular choice of coordinates, this method writes a subset of the components of Einstein's equation in a manifestly elliptic form, whilst the remaining components (which are not elliptic) are treated as constraints. The latter are dealt with by a careful imposition of boundary conditions. Even though this approach has been successfully used in a number of interesting problems it is not very robust in practice, and it is limited to co-homogeneity two metrics. See also \cite{Chesler:2013qla}.

A completely general method for solving Einstein's equation in static spacetimes was proposed in \cite{Headrick}. This method was further extended to the stationary case in \cite{TW} (see \cite{Wiseman:2011by,Dias:2015nua} for reviews). We will discuss this method in \S\ref{sec:reviewmethod}, but here we just outline the salient features that motivate the present work. Solving the time independent Einstein's equation numerically is not straightforward due to the underlying diffeomorphism invariance of the theory which implies that the equations do not have a well defined character. The main idea of \cite{Headrick,TW} is that instead of solving Einstein's equation \emph{after gauge fixing}, one solves the related ``harmonic'' Einstein's equation which has a manifest elliptic character and can therefore be readily solved using standard numerical methods for elliptic partial differential equations. Solving these harmonic  equations can be thought of as attempting to solve both the actual Einstein's equation \textit{together with} gauge conditions \textit{simultaneously} \footnote{This should be contrasted with the numerical solution of the time-dependent Einstein's equation in (generalised) harmonic gauge. In this case, one imposes the (generalised) harmonic gauge condition in the initial time slice and the gauge is preserved to the future due to the hyperbolic character of the equations.}. Solutions divide into two categories. Those that are solutions to the original Einstein's equation in a specific gauge (the `wanted' solutions), and those that are not.
In the case that the matter is a cosmological constant these latter solutions are Ricci solitons (see for example \cite{Topping}). 

For zero or negative cosmological constant matter ref. \cite{Lucietti} managed to rule out the existence of the unwanted Ricci solitons in static asymptotically flat, asymptotically Kaluza-Klein and asymptotically AdS spacetimes. However there is much interest in studying stationary but non-static black hole spacetimes and the method of \cite{Headrick,TW} has been successfully used in a number of cases (see \cite{Dias:2015nua} for a recent review) that are not covered by the result of \cite{Lucietti}. The existence of problematic unwanted Ricci soliton solutions has not been encountered in these applications, suggesting that their existence may be tightly constrained, and that the static results of \cite{Lucietti} may generalise to the stationary case as well. Understanding this generalisation is the purpose of this paper, and again for zero or negative cosmological constant we will indeed find the non-existence of stationary Ricci solitons that possess a ``$t$-$\phi$" reflection symmetry and which have a vanishing ``soliton charge'' -- a quantity we define from the asymptotics of the solution. The numerical application of this is that solving the harmonic formulation of Einstein's equation for such spacetimes, and imposing asymptotic boundary conditions to ensure vanishing soliton charge, then one is guaranteed that the only solutions found are indeed solutions of the original Einstein's equation.

 The rest of this paper is organised as follows. In \S\ref{sec:reviewmethod} we review ``$t$-$\phi$" reflection symmetry for spacetimes, and the harmonic formulation of the stationary Einstein's equation given in \cite{Headrick,TW}. In \S\ref{sec:nonexistence} we present our main result. After some preliminaries, and a brief review of the static non-existence result of \cite{Lucietti}, we define the soliton charge and show that when this vanishes then there exist no ``$t$-$\phi$" reflection symmetric Ricci soliton solutions  with zero or negative cosmological constant. In \S\ref{sec:application} we then discuss application of this result to the harmonic Einstein's equation for asymptotically flat and asymptotically AdS black hole spacetimes. Finally, in \S\ref{sec:discussion} we summarise our results and outline future research directions.  

%
\section{Review of numerical method for stationary vacuum solutions}
\label{sec:reviewmethod}
%

Suppose we wish to solve the $D$-dimensional pure gravity Einstein's equation with cosmological term for a stationary (i.e., time independent) spacetime,
\begin{eqnarray}
\label{eq:Einstein}
R_{\mu\nu} = \Lambda\, g_{\mu\nu} \; .
\end{eqnarray}
Here we will assume a vanishing or negative cosmological constant, so that $\Lambda \le 0$.
This system is not well posed as a p.d.e. problem due to the coordinate invariance. It was proposed in \cite{Headrick} that the DeTurck `trick' of Ricci flow \cite{DeTurck} is employed in order to turn \eqref{eq:Einstein} into a well-posed (elliptic) system. Here a vector $\xi^\mu$ is constructed as the difference of the usual (i.e., Levi-Civita) connection $\Gamma^\alpha_{~\mu\nu}$ of the spacetime metric $g$ and a fixed smooth reference connection $\bar{\Gamma}^\alpha_{~\mu\nu}$ as,
\begin{eqnarray}
\xi^\mu = g^{\rho\sigma}\left(\Gamma^\mu_{~\rho\sigma} - \bar{\Gamma}^\mu_{~\rho\sigma}\right) \; .
\end{eqnarray} 
This yields a globally defined one form $\xi_\mu = g_{\mu\nu}\,\xi^\nu$, from which we construct what we term the ``harmonic'' Einstein's equation,
\begin{eqnarray}
\label{eq:harmonic}
R^H_{\mu\nu} \equiv R_{\mu\nu} - \nabla_{(\mu} \xi_{\nu)} = \Lambda \, g_{\mu\nu} \; .
\end{eqnarray}
Now this equation has a principle part given by,
\begin{eqnarray}
R^H_{\mu\nu} =_{PP} - \frac{1}{2} g^{\alpha\beta} \partial_\alpha \partial_\beta g_{\mu\nu}
\end{eqnarray}
and hence the character of the harmonic Einstein's equation is just determined by the inverse metric $g^{\mu\nu}$.

Let us assume for now that we have a stationary spacetime with Killing vector $K = \partial / \partial t$ which is asymptotically timelike. Under certain conditions the rigidity theorem \cite{Hawking:1971vc,Hawking:1973uf,Hollands:2006rj,Moncrief:2008mr} guarantees that there will exist at least another rotational Killing field $\Phi=\partial / \partial\phi$ that commutes with $K$ \footnote{The rigidity theorem assumes that the spacetime is analytic, which is a rather strong assumption. See \cite{Ionescu:2007vk,Alexakis:2009gi,Alexakis:2013rya} for recent progress in removing this assumption from the actual proof of the theorem.}. We shall consider the situation where there exist $n\geq 1$ such commuting rotational Killing fields, $\Phi_\Lambda$,  $\Lambda=1,\ldots,n$. Furthermore, we assume that any horizons are bifurcate Killing horizons and are generated by $T = K+\Omega^\Lambda\,\Phi_\Lambda$, with non-vanishing surface gravity and angular velocities $\Omega^\Lambda$.  Let us denote by $\Psi$ the set of commuting Killing vector fields, $\Psi = \{K, \Phi_1,\ldots,\Phi_n\}$. If we further assume that the action of the isometry group is trivial \footnote{See \cite{Wald} for a precise definition.}, we may appeal to the recent result of \cite{Wald} according to which the spacetime metric has a ``$t$-$\phi$" reflection symmetric form, meaning it can be written in a block diagonal form as a trivial fibration over a base manifold $(\mathcal M,\hat h)$ \footnote{Ref. \cite{Wald} only proved this in the asymptotically flat case, but the generalisation to the AdS case should be straightforward}:
\begin{equation}
ds^2= G_{AB}(x)\,dy^A\,dy^B + \hat h_{ij}(x)\,dx^i\,dx^j\,
\label{eq:metric}
\end{equation}
where the indices $A,B,\ldots$ label the Killing directions and $i,j,\ldots$ are indices along the base $\mathcal M$. Here $G_{AB}$ is a matrix of smooth scalar functions on $\mathcal M$. The terminology of ``$t$-$\phi$" reflection symmetry refers to the invariance under $y^A \to - y^A$, which physically means the spacetime is invariant under simultaneously reversing time and the sense of rotation
of the black hole horizons. In the case of the Kerr metric (given explicitly in the later equations~\eqref{eq:Kerrfibre} and~\eqref{eq:Kerrbase}), where the Killing vectors are $\partial / \partial t$ and $\partial / \partial \phi$, indeed the spacetime is invariant under $t \to -t$, $\phi \to - \phi$, hence the ``$t$-$\phi$" terminology.

Since the spacetime is stationary then $\mathcal M$ has an asymptotic end where $\det{G_{AB}} < 0$. We have the fibre metric degenerate, so $\det{G_{AB}} = 0$, at base points corresponding to spacetime horizons, where the Killing vectors that generate the horizons becomes null, or axes of symmetry, where rotational Killing vectors degenerate.
Restricting consideration to the spacetime on, and exterior to any black hole horizons, then the interior of the base $\mathcal{M}$ is the region $\det{G_{AB}} < 0$ and its boundary $\partial \mathcal{M}$ is formed by the base points corresponding to horizons and axes of symmetry where $\det{G_{AB}} = 0$.
Thus $G_{AB}$ is a Lorentzian metric in the interior of $\mathcal{M}$ and at any interior point we may find a linear combination of the Killing vectors which is timelike. Since the full spacetime metric has Lorentzian signature then the base metric $\hat h$ must be Riemannian. 
Note that the full spacetime may have ergoregions, as we do not require the same linear combination of the $\Psi$ is everywhere timelike. 

For simplicity, in this paper we take the reference connection $\bar{\Gamma}^\alpha_{~\mu\nu}$ to be associated to a reference metric $\bar g$ on the same spacetime manifold of interest. Furthermore, we take  $\bar{g}$ to have the same block diagonal form as the spacetime metric $g$, with $(G, \hat h_{ij}) \to (\bar G,\bar{\hat h}_{ij})$, so that $\bar{G}_{AB}$ and $\bar{\hat h}$  are  smooth scalar functions and a smooth metric  over $\mathcal M$.
We impose that at any Killing horizon of the spacetime metric $g$, the reference metric also gives rise to a smooth Killing horizon with the same null generator and surface gravity $\kappa$ as that of the spacetime metric $g$. Similarly, we impose that all rotational Killing fields have the same periodicities.
Then the harmonic Einstein's equation~\eqref{eq:harmonic} consistently truncates to this stationary block diagonal form, forming a p.d.e. system that may be solved for the unknown $(G, \hat h_{ij})$. 

For a metric of the form \eqref{eq:metric} the only non-vanishing connection components are,
\begin{align}
\Gamma^A_{~Bi} & = \frac{1}{2}\,G^{AC}\partial_iG_{BC} \; , \quad
\Gamma^i_{~AB}  = -\frac{1}{2}\, \hat h^{ij}\partial_j G_{AB}\;,\quad
\Gamma^{k}_{~ij} = \hat\Gamma^k_{~ij}\;,
\end{align}
where $\hat\Gamma^k_{~ij}$ is the Christoffel symbol 
associated to
the  base metric $\hat h$. An important consequence is that the Killing fields are orthogonal to the vector $\xi$ as,
\begin{eqnarray}
\label{eq:xiorthog}
\xi_A = 0 \quad \implies \quad \Psi^\mu \xi_\mu = 0 \;.
\end{eqnarray}

Since we have adapted coordinates to our Killing symmetries, so $\frac{\partial }{ \partial y^A } g_{\mu\nu} = 0$, then the principle symbol of the harmonic Einstein's equation becomes,
\begin{eqnarray}
R^H_{\mu\nu} =_{PP} - \frac{1}{2} \hat{h}^{ij} \partial_i \partial_j g_{\mu\nu}
\end{eqnarray}
as $g^{ij} = \hat{h}^{ij}$ in our ansatz, where $\hat{h}^{ij}$ is the inverse metric to $\hat{h}_{ij}$. Now since we have taken $(\mathcal{M}, \hat{h})$ to be a smooth Riemannian manifold then, subject to appropriate asymptotic conditions and boundary conditions that ensure regular Killing horizons and axes of symmetry in the full spacetime, we see that the system of equations we have to solve is of elliptic character, and hence it can be tackled numerically using standard elliptic methods.

In our ``$t$-$\phi$" reflection symmetric case the base $\hat{h}$ is constrained to be Riemannian. More generally for stationary spacetimes as above where off-diagonal $dy^A dx^i$ terms are permitted in~\eqref{eq:metric} it is then not automatic that $\hat{h}$ is Riemannian everywhere. In particular for the flowing funnels \cite{Fischetti:2012vt} and the plasma flows \cite{Figueras:2012rb} it is not.  In such cases the horizons are not Killing and the system of equations is no longer elliptic, and hence it is not clear that the system is well posed anymore. Thus we will not consider this situation further here, although note that it would be interesting to develop the theory behind these more complicated solutions without Killing horizons. Note that in the examples of \cite{Fischetti:2012vt,Figueras:2012rb}, the spacetime is not ``$t$-$\phi$" symmetric since the black hole moves in a preferred direction with respect to stationary observers at infinity and this direction is not a symmetry of the spacetime.

An important question is whether a solution to the harmonic Einstein's equation \eqref{eq:harmonic} will give us a solution to the original Einstein's equation \eqref{eq:Einstein}. Even if $\xi^\mu \to 0$ asymptotically (see Section \ref{sec:application}) there may be other solutions with non-trivial $\xi^\mu$. These are to be regarded as spurious solutions since they do not solve the original Einstein's equation. Technically such spurious solutions, which solve \eqref{eq:harmonic} for some non-trivial vector field $\xi_\mu$, are called Ricci solitons, termed steady in the case $\Lambda = 0$ and expanding for $\Lambda < 0$. In practice one can simply check whether a solution found is a good Einstein metric or an unwanted Ricci soliton, and in the latter case one simply has to try again to find a different solution. The utility of this harmonic approach in a practical numerical setting then hinges on how many Einstein metrics  there are relative to the unwanted Ricci solitons, and how easily these are found amongst a potential morass of unwanted solutions. For this practical reason it is important to gain control over the existence of the unwanted Ricci solitons.

In certain instances one can prove analytically that no Ricci soliton solutions exist, and then every solution of the harmonic system is one of physical relevance as it is Einstein. For the closed Riemannian case it has long been known that no Ricci solitons exist \cite{Bourguignon,Perelman}. Note that we are not considering the case of positive cosmological constant as even in the closed Riemannian case explicit examples of Ricci solitons are known to exist (for example \cite{Koiso,Cao}) and presumably continue to do so in the Lorentzian setting. More recently it was shown that starting in the Lorentzian static setting with asymptotic regions and horizons, it remains true that solitons cannot exist \cite{Lucietti}.  The purpose of this paper is to show how to extend such a result to the stationary case. As we shall see, this is rather more subtle than in the static situation, and we will require ``$t$-$\phi$" reflection symmetry and an asymptotic condition of vanishing soliton charge in order to obtain a non-existence result.

The non-existence of Ricci solitons is clearly an optimal situation as any numerical solution found should be Einstein, and there is no concern of having to filter out (potentially many) unwanted solutions. Indeed, in such a case, the vector $\xi^\mu$ becomes a useful numerical tool since it is guaranteed to vanish on a continuum solution, and so its  residual value on a numerical solution gives a useful measure of the solution's accuracy. For example, the maximum value of its norm has been used extensively in previous work \cite{Lucietti} giving an excellent global measure of accuracy, and also can be used to check numerical convergence.
%

%
\subsubsection{Horizons and axes of symmetry}
\label{subsec:horBCs}
%

In this subsection we briefly review the boundary conditions that must be imposed  on horizons and axes of symmetries to ensure the smoothness of the spacetime geometry. The general case has been discussed in detail in \cite{TW}, and our line element \eqref{eq:metric} corresponds to the so called ``reduced symmetry" case discussed in that reference.

Consider a Killing horizon generated by the Killing vector $T = K + \Omega^\Lambda \Phi_\Lambda$. We may choose linear combinations of the commuting Killing vectors, $\partial/\partial y'^A = C_A^{~B} \partial/\partial y^B$, so that $T = \partial/\partial y'^1$ and we denote the remaining Killing vectors as $\partial/\partial y'^{\bar{A}}$. The horizon forms a boundary to the base $\mathcal M$, where $\partial/\partial y'^1$ degenerates, and locally we may take Gaussian normal coordinates to the horizon such that,
\begin{equation}
ds^2_{\mathcal M} = dr^2 + \hat h_{\bar i\bar j}\,dx^{\bar i}\,dx^{\bar j}\,,
\label{eq:metricHorBase}
\end{equation}
and the horizon is located at $r = 0$.
Then \cite{TW} showed that the metric on the fibres in \eqref{eq:metric} can be locally written as
\begin{equation}
G_{AB} dy^A dy^B = - r^2 \left( \kappa^2 + r^2 f \right)  \, (dy'^1)^2 + r^2\,f_{\bar{A}}\,dy'^1\,dy'^{\bar{A}} + G'_{\bar A \bar B}\,dy'^{\bar A}\,dy'^{\bar B}  \,.
\label{eq:metricHorFibre}
\end{equation}
The constraint that this spacetime metric yields a regular stationary horizon implies that $\kappa$ is a constant and all the component functions of the base and fibre metrics, $\hat h_{\bar i \bar j}$, $f$, $f_{\bar A}$ and $G'_{\bar A \bar B}$, in this local chart must be smooth functions of $(r, x^{\bar i})$ and in addition they must be  \emph{even} functions in $r$. Hence these components are all smooth functions of $x^{\bar i}$ and $r^2$. The constant $\kappa$ gives the surface gravity of the Killing horizon with respect to the Killing vector $T$, and together with the angular velocities $\Omega^\Lambda$, may be regarded as boundary data for the elliptic problem.

As discussed in  \cite{TW}, we emphasise that the conditions above do not depend on the spacetime being a solution to the Einstein's equation, but arise simply by demanding a regular Lorentzian Killing horizon. Regularity is demonstrated simply in analogy with polar coordinates where the analog `Cartesian' coordinates are $\sigma, \rho$,
\begin{equation}
\label{eq:horizcoord}
\sigma = r \cosh{ \kappa \, y'^1} \; , \quad \rho = r \sinh{ \kappa \, y'^1} \; .
\end{equation}
Taking the same chart on the base, these boundary conditions also translate to the reference metric, which must take the same form with metric functions $\bar{ \hat{h}}_{\bar i \bar j}$, $\bar f$, $\bar f_{\bar A}$ and $\bar G'_{\bar A \bar B}$ that are smooth in $(r, x^{\bar i})$ and even in $r$, and where the constant $\kappa$ must be taken to be the same as that for the metric so that the harmonic Einstein's equation is regular.

Of relevance for us here is that in order to respect the Killing symmetries of the spacetime and be smooth, a scalar field $\psi$ must also be a smooth function of $(r^2, x^{\bar i})$. Furthermore a general covector field $v$ that is smooth, respects the symmetries of the spacetime, and is orthogonal to the Killing directions must take the form,
\begin{equation}
\label{eq:vecboundary}
v = r \, v_r \, dr + v_{\bar i}\, dx^{\bar i}
\end{equation}
where the component functions $v_r$ and $v_{\bar i}$ are smooth in $(r^2, x^{\bar i})$. In particular, provided the reference metric is chosen to be regular as discussed, then the covector $\xi$ has this behaviour. Note that regularity does not require $\xi$ to vanish at a horizon.

The boundary conditions that ensure the smoothness of the geometry at an axis of symmetry are analogous to the horizon ones discussed above. The base metric can again locally be taken to have Gaussian normal form about the axis at $r = 0$. We now choose linear combinations of the commuting Killing vectors, $\partial/\partial y''^A = {C'}_A^{~B} \partial/\partial y^B$, so that $R = \partial/\partial y''^1$ degenerates at $r=0$ and has period $2 \pi$ and we denote the remaining Killing vectors as $\partial/\partial y''^{\bar{A}}$. Then the fibre metric is,
\begin{equation}
G_{AB} dy^A dy^B  =  r^2 \left( 1 + r^2 f \right)  \, (dy''^1)^2 + r^2\,f_{\bar{A}}\,dy''^1\,dy''^{\bar{A}} + G''_{\bar A \bar B}\,dy''^{\bar A}\,dy''^{\bar B}  \,,
\label{eq:metricAxisFibre}
\end{equation}
and the metric functions $\hat h_{\bar i \bar j}$, $f$, $f_{\bar A}$ and $G''_{\bar A \bar B}$ are smooth in $(r^2, x^{\bar i})$. Then $(r,y''^1)$ are polar coordinates for the axis, and we may manifest smoothness of the full spacetime by moving to the usual Cartesian coordinates, $X = r \cos{y''^1}$, $Y = r \sin{y''^1}$. The same considerations as for a horizon apply to the behaviour of a scalar or vector (orthogonal to the Killing directions) that preserve the Killing symmetries. The generalisation required when two boundaries meet is straightforward, and discussed explicitly in \cite{TW}. 

%
\section{Non-existence of stationary solitons}
\label{sec:nonexistence}
%

Let us now state our result for non-existence of steady ($\Lambda = 0$) and expanding ($\Lambda < 0$) stationary Ricci solitons with ``$t$-$\phi$" reflection symmetry. We shall then go on to provide its justification and, in the next section, consider its application to our motivating numerical problem.\\

\noindent
{\bf Setting:}\\
Consider a $D$-dimensional stationary spacetime on, and exterior to any horizons, a set of commuting Killing vector fields $\Psi= \{K, \Phi_1,\ldots,\Phi_n\}$, one of which, $K$, is asymptotically timelike, and a one-form $v$, symmetric under the Killing vectors of $\Psi$. We assume the following; 
\begin{itemize}
\item
Denote connected horizon components as $\{ \mathcal{H}_A \}$. A horizon $\mathcal{H}_A$ is a smooth bifurcate Killing horizon generated by $T=K+\Omega^\Lambda_{(A)}\,\Phi_\Lambda$, with a non-vanishing surface gravity $\kappa_{(A)}$. The spacetime and the one-form $v$ are smooth on, and exterior to the horizons.
\item
On, and exterior to any horizons, the spacetime and one-form can be written in ``$t$-$\phi$" reflection symmetric form as a fibration over a smooth Riemannian base $(\mathcal M, \hat{h})$, with boundaries $\partial \mathcal{M}$ corresponding to horizons and axes of symmetry,
\begin{align}
\label{eq:spacetime}
ds^2 = g_{\mu\nu} dx^\mu dx^\nu= G_{AB}(x)\,dy^A\,dy^B + \hat{h}_{ij}(x) dx^i dx^j \; , \quad v = \hat{v}_i(x) dx^i \; , \quad \det G \le 0 \; ,
\end{align}
with ``$t$-$\phi$" reflection symmetry of the one-form implying it is orthogonal to the Killing vectors of $\Psi$.
\item
 There are no boundaries to the spacetime but there may be asymptotic ends. Let $\hat{R}(x^i)$ be a function such that it is finite in the interior of the spacetime and $\hat{R} \to \infty$ in any asymptotic end. We define the following two scalars $\phi$ and $\omega$, and also a ``soliton charge'' $Q$ as,
 \begin{align}
\label{eq:defn}
\phi & = v_\mu v^\mu\,, \nonumber \\
\omega & = \phi + \nabla^\mu v_\mu\, , \nonumber \\
Q & = \lim_{\hat{R} \to \infty} \int_{\hat{R}} {\hat{dS}}^i \sqrt{|G|}\, \hat{v}_i 
\end{align}
with $\nabla_\mu$ the covariant derivative of the full spacetime. We demand the asymptotic behaviour is such that $\phi , \omega \to 0$ sufficiently fast that the integrals $\int_{\mathcal M} dx\,\sqrt{\hat h}\,\sqrt{|G|}\, \phi$, $\int_{\mathcal M} dx\,\sqrt{\hat h}\,\sqrt{|G|}\, \omega$ 
do not diverge.
 \end{itemize}

\noindent
{\bf Claim:}\\
For ``$t$-$\phi$" reflection symmetric spacetimes and one-forms $v$ as above, then a solution of the steady/expanding Ricci soliton equation,
\begin{align}
\label{eq:solitoneq}
R_{\mu\nu} - \nabla_{(\mu} v_{\nu)} = \Lambda\, g_{\mu\nu}
\end{align}
with $\Lambda \le 0$, has non-vanishing one-form $v$ if and only if $Q < 0$. In particular if $Q =0$ the solution must be an Einstein metric, and $Q > 0$ is not possible. 
\\

We should remark here that $Q$ is defined only for stationary spacetimes of the above form, and thus is not an actual dynamically conserved charge in the usual sense. The term ``charge'' is used to reflect the dependence of $Q$ only on the asymptotics of the solution.
\\

As a corollary, taking $v = \xi$ to be a covector field constructed from a ``$t$-$\phi$" reflection symmetric reference connection in the way of DeTurck (so it will be orthogonal to the Killing vectors of $\Psi$), then together with asymptotic conditions that ensure $Q = 0$, this implies non-existence of spurious solutions to the harmonic Einstein system for zero or negative cosmological constant matter. We now proceed to give the argument for this claim.

%
\subsection{Preliminaries}
%

In this subsection we derive some preliminary results that will be useful in the subsequent discussion. 
For a solution of the soliton equation~\eqref{eq:solitoneq} above, the contracted Bianchi identity implies,
\begin{align}
\label{eq:Bianchi}
\nabla^2 v_\mu + R_{\mu}^{~~\nu} v_\nu = 0 \; .
\end{align}
Then contracting the above Bianchi vector equation \eqref{eq:Bianchi} with $v^\mu$, and by taking its divergence, we  obtain the following two scalar equations,
\begin{align}
\label{eq:phieqn}
\nabla^2 \phi + v^\mu \partial_\mu \phi & = - 2 \,\Lambda \, \phi  + 2\,( \nabla_{\mu} v_\nu) (\nabla^{\mu} v^\nu) \,, \\
\label{eq:omegaeqn}
\nabla^2 \omega + v^\mu  \partial_\mu  \omega  & = - 2 \, \Lambda \, \omega + \frac{1}{2} F_{\mu\nu} F^{\mu\nu} \,,
\end{align}
where we have defined the 2-form $F$ from the antisymmetric part of $\nabla v$ as,
\begin{align}
F_{\mu\nu} = 2\, \partial_{[\mu} v_{\nu]} \, .
\end{align}
These are the basic equations that we will use to show non-existence of solutions with $v \ne 0$ with appropriate asymptotics.

Recall that given the assumptions above, the metric $\hat{h}$ defines a Riemannian base $(\mathcal{M}, \hat{h})$. Firstly, since $v_A = 0$, and $\partial_A {v}_i = 0$, then the covector $v$  reduces to a covector $\hat{v} = \hat{v}_i(x) dx^i$ on the base. Similarly, since $\partial_A \phi = \partial_A \omega = 0$, these scalars  reduce to functions over this base.
Then,
\begin{align}
\label{eq:defn2}
\phi & = \hat{v}^i \hat{v}_i\,, \nonumber \\
\omega & =  \hat{v}^i \hat{v}_i + \hat{\nabla}^i \hat{v}_i + \hat v^i\hat J_i\,,\quad \hat J_i = \frac{1}{2}\,G^{AB}\,\partial_iG_{AB} = \partial_i \ln \sqrt{ | G | } \, ,
\end{align}
where these expressions are written covariantly with respect to the base -- $\hat{\nabla}$ is the covariant derivative compatible with $\hat{h}$, and indices are raised/lowered using this base metric. Since $\hat{h}$ is Riemannian, then,
\begin{align}
\phi & = \hat{h}^{ij} \hat{v}_i \hat{v}_j \ge 0 \; .
\end{align}
Again due to $v_A = 0$ and $\partial_A {v}_i = 0$,  the 2-form $F_{\mu\nu}$ has components,
\begin{align}
F_{AB} = 0 \; , \quad F_{Ai} =  F_{iA}  = 0 \; , \quad F_{ij} = \hat{F}_{ij} = 2\, \partial_{[i} \hat{v}_{j]}\,,
\end{align}
and so reduces to the antisymmetric 2-form $\hat{F}_{ij}$ on the base.  A consequence of this, that will be of relevance to the argument which follows, is,
\begin{align}
F_{\mu\nu} F^{\mu\nu} = \hat{F}_{ij} \hat{F}^{ij} \ge 0 \, .
\end{align}
Note however that,
\begin{equation}
\label{eq:dvsqr}
{(\nabla_{\mu} v_\nu) (\nabla^{\mu} v^\nu)}  =  \frac{1}{4} | \hat{F} |^2  +  \frac{1}{4} |  \hat{H} |^2  + \frac{1}{4}\,G^{AB}\,G^{CD}\,\left(\hat v^i\partial_iG_{AC}\right)\left(\hat v^j\partial_j G_{BD} \right)\, ,
\end{equation}
where we have defined, $\hat{H}_{ij}  = 2\, \hat{\nabla}_{(i} \hat{v}_{j)}$, and the last term in \eqref{eq:dvsqr} in general will not have a definite sign (since $G_{AB}$ is not positive definite) and this will be important later. 
Lastly the differential operator $\nabla^2$ acting on a scalar reduces to,
\begin{align}
\label{eq:reduc1}
\nabla^2 =  \hat{\nabla}^2   +  \hat{J}^i \partial_i 
\end{align}
over the base. Note that while the metric $\hat{h}$ is regular at any horizon or axis boundary in the base, the vector  $\hat{J}_i = \partial_i \ln \sqrt{ | G | }$ is not since $\sqrt{|G|} \to 0$ there. This singular behaviour is due to axes and horizons being coordinate singularities when reducing on the isometry directions, and is something we will discuss later.

%
\subsection{Review of previous argument for static case}
\label{sec:reviewstatic}
%

We now review the previous arguments in \cite{Lucietti} for non-existence of static solitons with $\Lambda \le 0$. 
Since we are considering static solutions to \eqref{eq:harmonic}, and $v^\mu$ is compatible with staticity, it is convenient to continue to Euclidean time $\tau = i t$ so the static Lorentzian spacetime becomes a Riemannian solution to~\eqref{eq:solitoneq}. Then, in this Riemannian signature we see $\phi = v^\mu v_\mu \ge 0$ and $(\nabla_\mu v_\nu) (\nabla^\mu v^\nu) \ge 0$ and so for $\Lambda \le 0$  equation \eqref{eq:phieqn} then implies,
\begin{align}
\label{eq:maxphieqn}
\nabla^2 \phi + v^\mu \partial_\mu \phi & \ge 0 \,,
\end{align}
where $\nabla^2$ is an elliptic operator. Hence the maximum principle then states that if $\phi$ achieves its maximum in the interior of the geometry, it must be constant everywhere.

However this leaves the possibility that $\phi$ might attain a maximum at a horizon or asymptotically. Consider a particular horizon component with its surface gravity $\kappa$. Taking $\tau \sim \tau + 2 \pi \kappa$ removes this horizon as a boundary of the Riemannian space, with the geometry being smooth and simply the set of points where the isometry generated by $\partial/\partial \tau$ has fixed points.  Now we apply the maximum principle in the vicinity of this horizon, with appropriately periodic $\tau$. Since the horizon points no longer form a boundary of the geometry, but are interior points, a non-constant $\phi$ cannot attain a maximum at them. However, since we are considering static vector fields, the vector does not depend on the Euclidean time coordinate. Thus, the fact that no maximum is allowed at that horizon is independent of whether $\tau$ is chosen to be periodic. Likewise, it is independent of whether we work with Euclidean or Lorentzian time. For each horizon component we may make the same argument (with appropriately chosen period given in terms of the horizon surface gravity) and thus conclude that $\phi$ is either constant, or cannot attain a maximum in the interior of the spacetime or at any Killing horizon. Hence any maximum value must be attained asymptotically.

Then provided we have asymptotic conditions such that $\phi \to 0$ asymptotically,  this maximum principle implies that $\phi$ can have no positive value. However by construction $\phi \ge 0$, and so this implies $\phi$ must vanish everywhere. Since $\phi$ is the norm of $v^\mu$, and the geometry is Riemannian, then the vanishing of $\phi$ implies the vanishing of $v^\mu$ everywhere, and hence the non-existence of a static soliton. As discussed in \cite{Lucietti}, in the numerical setting where we solve the harmonic Einstein's equation, then for appropriately chosen reference connections one indeed ensures $\phi \to 0$ for both the asymptotically flat ($\Lambda =0$) and AdS ($\Lambda < 0$) cases. Hence in this static vacuum context the method is guaranteed to find only the desired Einstein solutions.

The key tool we have employed in the static argument is the ability to continue a solution of~\eqref{eq:solitoneq} to Euclidean time. In the stationary setting we can no longer turn the Lorentzian problem into a real Riemannian one simply by taking $t = i \tau$, due to the `off-diagonal' time-space components in the metric which become imaginary. Whilst for an analytic solution, one can expect to make a continuation of the parameters of the solution (e.g., angular momenta) to obtain a real Riemannian geometry, this is irrelevant here where we are not assuming analytic properties of solutions, and furthermore want a method to find solutions which we do not know. Constructing a solution numerically does not generally allow one access to its analytic continuation. For these reasons the stationary case cannot simply be tackled by the method of continuation.

More importantly the first key step in the static case no longer follows. Looking at the norm ${(\nabla_{\mu} v_\nu) (\nabla^{\mu} v^\nu)}$ we see from \eqref{eq:dvsqr} that in general the last term depends on the matrix of scalar fields $G_{AB}$ and does not have a definite sign. One might wonder whether some identity could ensure the collection of terms is always non-negative. This is simple to contradict with an explicit example.
Consider taking the stationary spacetime and vector $v$ as,
\begin{align}
ds^2 = - \left( dt +  \psi(x) \,  dy \right)^2 + \delta_{ij} dx^i dx^j \; , \quad v_\mu = \partial_\mu V(x)
\end{align}
in local coordinates $x^i = (x, y, z^a)$ so $a = 1, \ldots , D-2$, for functions $\psi, V$  which depend only on the $x$ coordinate. This is of ``$t$-$\phi$'' symmetric form. Since $\hat{v} = d V$, then $\hat{F} = 0$, and one finds that the last term in \eqref{eq:dvsqr} may dominate the positive $\hat{H}^2$ term. Explicitly,
\begin{align}
{(\nabla_{\mu} v_\nu)(\nabla^{\mu} v^\nu)} = - \frac{1}{2} \left( \psi' V' \right)^2 + ( V'' )^2
\end{align}
where the prime indicates differentiation w.r.t. $x$. We clearly see that this may be negative or positive for appropriate choices of $V$ and $\psi$. Since ${(\nabla_\mu v_\nu)(\nabla^\mu v^\nu)}$ may be negative in general, the approach in the stationary case cannot simply follow that of the static case, using simply equation \eqref{eq:phieqn}.

%
\subsection{Argument for the stationary case}
\label{sec:newarguement}
%

In this subsection we will provide a new argument to establish the non-existence of Ricci solitons that applies to the stationary case. This new argument has two components. Firstly, based on the second scalar equation \eqref{eq:omegaeqn} that follows from the Bianchi identity, we will derive a maximum principle which implies that if $\omega$ is not constant then it cannot attain a non-negative maximum value in the interior of the base $\mathcal{M}$ or its boundaries $\partial \mathcal{M}$. Secondly, we use the assumed asymptotic behaviour to show this maximum principle implies $\omega\leq 0$, and further that $v$ must vanish.

For clarity of presentation we will begin with the second part which is straightforward to show, and then give the details of the maximum principle which is somewhat more technical.

%
\subsubsection*{Part 1: $\omega \leq 0$ $\implies$ $v^\mu = 0$}
\label{subsec:part1}
%

Let us assume our stated maximum principle, that $\omega$ is either constant or, if not, cannot attain a non-negative maximum value in the interior of the base $\mathcal{M}$ nor at its boundaries $\partial \mathcal{M}$ associated to axes of symmetry or horizons. Then, any non-negative maximum value will be attained asymptotically. However, our boundary conditions imply that $\omega \to 0$ asymptotically, and hence our maximum principle implies that $\omega$ cannot be positive anywhere. Thus we conclude that over the base $\mathcal M$ we have,
\begin{align}
 \omega & \le 0 \, .
\end{align}
We now integrate the function $\sqrt{|G|}\,\omega$ over $\mathcal{M}$ to get,
\begin{align}
\int_\mathcal{M} dx \,\sqrt{\hat{h}}\, \sqrt{|G|} \,\omega & = \int_\mathcal{M} dx\, \sqrt{\hat{h}}\, \sqrt{|G|} \left( \phi + \hat{\nabla}^i \hat{v}_i +  \hat v^i \hat J_i \right) \le 0\,.
\end{align}
Now rearranging and integrating by parts we find,
\begin{align}
\label{eq:bdryterm}
\int_\mathcal{M} dx \,\sqrt{\hat{h}}\, \sqrt{|G|} \phi \le - \int_\mathcal{M} dx \,\sqrt{\hat{h}} \,\hat{\nabla}^i \left(    \sqrt{|G|} \hat{v}_i \right) = - \int_\mathcal{\partial M} dS^i   \sqrt{|G|}\, \hat{v}_i - \lim_{\hat{R} \to \infty} \int_{\hat{R}} dS^i   \sqrt{|G|}\, \hat{v}_i 
\end{align}
 where 
$\hat{R}$ is the function defined above equation \eqref{eq:defn}, and $dS^i$ in the first integral is the outer directed volume element for $\partial \mathcal{M}$, and in the second is the outer directed (i.e., towards larger $\hat{R}$) volume element for a constant $\hat{R}$ hypersurface. Note that our asymptotic conditions have assumed that these integrals do not diverge.

Now from equation  \eqref{eq:vecboundary} we have that $\hat{n}^i \hat{v}_i = 0$ on the boundary $\partial \mathcal{M}$ corresponding to horizons, and similarly for the parts of $\partial \mathcal{M}$ due to axes of symmetry. Hence the first surface term on the right-hand side above vanishes, and the second is given in terms of the soliton charge, so we find,
\begin{align}
\label{eq:bdryterm}
  Q \le - \int_\mathcal{M} dx \,\sqrt{\hat{h}}\, \sqrt{|G|} \phi \, .
\end{align}
Now suppose a soliton solution exists which is not an Einstein solution, so $\hat{v}_i \ne 0$. Recalling that $\phi \ge 0$, with equality only where $\hat{v}_i$ vanishes, then this implies $Q < 0$. Non-vanishing $Q$ implies non-vanishing $\hat{v}_i$ and hence a soliton solution which is not Einstein, and thus $Q > 0$ is not possible. We also see that if $Q = 0$, the above implies that $\phi = 0$ and so $\hat{v}_i=0$ everywhere, and hence the solution is not a soliton, but is an Einstein metric. Hence we find the statements in our {\bf Claim}. Now to complete the argument we derive the maximum principle for $\omega$.

%
\subsubsection*{Part 2: The maximum principle for $\omega$}
\label{subsec:part2}
%

As we have discussed, ${(\nabla_\mu v_\nu)(\nabla^\mu v^\nu)}$ does not have definite sign in the stationary but non-static setting, and hence equation \eqref{eq:maxphieqn} does not generalise to this case. However the square of $F_{\mu\nu}$ does, being non-negative, and hence the scalar equation \eqref{eq:omegaeqn} yields the bound,
\begin{align}
\label{eq:ineqstationary}
\nabla^2 \omega + v^\mu  \partial_\mu  \omega  & \ge - 2 \, \Lambda \, \omega \, .
\end{align}
The equation above is so far applied to a Lorentzian spacetime, and thus we have to understand why a maximum principle (which requires an elliptic operator) can result. Due to the invariance on the Killing directions we may consider the equation reduced to the base $(\mathcal{M}, \hat{h})$ where we obtain,
\begin{align}
\label{eq:ineqstationary2}
\left( \hat{\nabla}^2  + \hat{J}^i \partial_i + \hat{v}^i \partial_i \right) \omega  & \ge - 2 \, \Lambda \, \omega \; .
\end{align}
Now since $\hat{h}_{ij}$ is Riemannian, this is an elliptic operator, and as $\Lambda \le 0$, this directly yields the following maximum principle -- a non-constant $\omega$ may not attain a non-negative maximum in the interior of the base $\mathcal{M}$. Any such maximum must occur at a boundary or asymptotic region of $\mathcal{M}$.

However, as we have discussed, $\mathcal{M}$ has boundaries, namely horizons and axes of symmetry in the full spacetime, and so we have to understand how to rule out putative non-negative maxima at such boundaries. These are coordinate singularities due to reducing on the Killing isometry directions and, as a consequence of this, the vector $\hat{J}_i$ is singular there. In the static case (where we didn't reduce on rotational isometry directions, so did not have axes to consider) we used the trick of continuing to periodic Euclidean time to remove horizons as boundaries and obtain a regular elliptic operator when we worked in the full Riemannian space. Here we cannot perform the Euclidean continuation in stationary spacetimes, but can use a related idea to proceed.

Consider a point $p$ in the full spacetime that lies on an axis of symmetry or horizon (and hence reduces to a point $\hat{p}$ in the boundary $\partial \mathcal{M}$ of the base). We can find a spacetime neighbourhood $\Omega_p$ of this point $p$ where we can write the spacetime metric as,
\begin{align}
\label{eq:spacetime}
ds^2 = - N(X)\, d\bar{t}^2 + 2\,A_a(X)\, d\bar{t}\, dX^a + \bar{g}_{ab}(X)\, dX^a \, dX^b  \; , \quad v = \bar{v}_a(X)\, dX^a \; , \quad N(X) \ge 0  \, ,
\end{align}
and all metric components are $\bar{t}$ independent, with $\partial / \partial \bar{t}$ being a linear combination of the commuting Killing vectors $\Psi$. Then $\bar{g}_{ab}$ is the spatial geometry of a constant $\bar{t}$ section of the spacetime, and is isometric under the other Killing vectors $\Psi$ than $\partial / \partial \bar{t}$. Since the full spacetime should be regular, this requires this spatial geometry also to be regular. Thus $\bar{g}_{ab}$ must be a smooth Riemannian geometry. The coordinates $X^a$ are formed from the base coordinates, $x^i$ and combinations of the fibre coordinates $y^A$. While written in the original ``$t$-$\phi$'' symmetric form this geometry appears singular at axes of symmetry, Cartesian coordinates may always be chosen to manifest its regularity explicitly.

If $p$ does not lie in a horizon then we may choose the linear combination $\partial / \partial \bar{t}$ and neighbourhood $\Omega_p$ so that $\partial/\partial \bar{t}$ is time-like over $\Omega_p$, with $N(X) > 0$. 
On the other hand if $p$ lies in a horizon then we may choose $\Omega_p$ so that $N = 0$ at the points in $\Omega_p$ which lie in this horizon and $\partial / \partial \bar{t}$ is its null generator, and $\partial/\partial \bar{t}$ is time-like elsewhere in $\Omega_p$. 
Then the line element~\eqref{eq:spacetime} is written in co-rotating coordinates.

We emphasise that the metric in equation \eqref{eq:spacetime} only applies over the neighbourhood $\Omega_p$ of $p$, and hence $\partial / \partial \bar{t}$ is not generally the asymptotic timelike Killing vector due to the presence of ergoregions. We cannot globally work with the spacetime in the form \eqref{eq:spacetime} and expect to have a Riemannian $\bar{g}_{ab}$.  However, in a small enough neighbourhood of any spacetime point we must always be able to find a Killing vector $\partial / \partial \bar{t}$ which is timelike away from points on horizons, and null on such points, so that the spatial sections with metric $\bar{g}_{ab}$ are Riemannian.

Let us denote the spatial slice of the spacetime neighbourhood $\Omega_p$ as $\bar{\Omega}$, and consider the Riemannian space $(\bar{\Omega}, \bar{g})$. An advantage of this space $(\bar{\Omega}, \bar{g})$ over the base $(\mathcal{M}, \hat{h})$ is that the former has no boundaries associated to axes of symmetry in the base, since the full spacetime does not, and we have not reduced on these spatial isometry directions in considering $\bar{\Omega}$. However, it still retains boundaries at horizons if $p$ lies in one.

Now consider our equation~\eqref{eq:ineqstationary} decomposed over the Riemannian base $(\bar{\Omega}, \bar{g})$. Then recalling that $\frac{\partial}{\partial \bar{t}} \omega = 0$, and that the covector $v$ is orthogonal to the Killing directions, and hence reduces to a covector $v = \bar{v} = \bar{v}_a dX^a$ over $(\bar{\Omega}, \bar{g})$, then,
we obtain the elliptic equation,
\begin{align}
\label{eq:ineqstationary2}
\left( \bar{\nabla}^2  + \bar{J}^a \partial_a+ \bar{v}^a  \partial_a  \right) \omega  & \ge - 2 \, \Lambda \, \omega \, , \qquad \bar{J}_a = \frac{1}{2} \partial_a \log\left( N + \bar{g}^{ab} A_a A_b \right) \; ,
\end{align}
where $\bar{\nabla}$ is the covariant derivative of $\bar{g}$ and indices are raised/lowered with respect to that metric.
Provided $p$ is not contained in a horizon, so $N > 0$ over $\bar{\Omega}$, the vectors controlling the single derivative terms are regular, and hence as $\bar{g}$ is smooth and Riemannian and $\Lambda \le 0$ the maximum principle for $\omega$ can be applied over $\bar{\Omega}$. Thus we see the maximum principle also applies at points corresponding to axes of symmetry. However we cannot make the argument at a horizon, as the covector $\bar{J}_a$ is still singular there. So we have learned that $\omega$ is either constant, or if not, cannot obtain a non-negative maximum outside a horizon.

Now we have to rule out such a maximum also at horizon points. Motivated by the static case, we consider a new Riemannian problem where we can remove these horizons boundaries. We consider the Riemannian metric built on the same base $(\bar{\Omega}, \bar{g})$ using the same function $N(X)$, but taking a spacelike `time' coordinate $\bar{\tau}$ and dropping `off-diagonal' time-space terms;
\begin{align}
\label{eq:aux}
ds^2_{(aux)} = g^{(aux)}_{\mu\nu} dx^\mu dx^\nu = N(X) \,d\bar{\tau}^2 + \bar{g}_{ab}(X) \, dX^a \, dX^b  \,.
\end{align}
Due to dropping these off-diagonal terms, we emphasise that this is not a Euclidean continuation of \eqref{eq:spacetime}, but should be thought of as an auxiliary Riemannian space. 

Suppose there is a Killing horizon with surface gravity $\kappa$ in ${\Omega}_p$, and hence a boundary component $\partial \bar{\Omega}_{\mathcal{H}}$ of $\bar{\Omega}$. Then provided ${\Omega}_p$ is taken sufficiently small we can choose the auxiliary metric over $\bar{\Omega}$ to have normal form to the the boundary $\partial \bar{\Omega}_{\mathcal{H}}$. Then our horizon boundary conditions, given in section \S\ref{subsec:horBCs}, imply that taking coordinates $X^a = ( r, X^{\bar{a}} )$ where $r$ is the normal coordinate, then the spacetime takes the form,
\begin{align}
\label{eq:spacetimenormal}
ds^2 = - r^2 \left( \kappa^2 + r^2 f(r,X^{\bar{a}}) \right) d\bar{t}^2 + r^2\, f_a \,d\bar{t} \,dX^{\bar{a}} + dr^2 + \bar{g}_{\bar{a}\bar{b}}(r,X^{\bar{a}})\, dX^{\bar{a}} \,dX^{\bar{b}}  
\end{align}
and so the auxiliary Riemannian space is,
\begin{align}
\label{eq:auxnormal}
ds^2_{(aux)} = r^2 \left( \kappa^2 + r^2 f(r,X^{\bar{a}}) \right) d\bar{\tau}^2 + dr^2 + \bar{g}_{\bar{a}\bar{b}}(r,X^{\bar{a}}) \, dX^{\bar{a}}\, dX^{\bar{b}}  
\end{align}
where the components $f$, $f_a$ and $\bar{g}_{\bar{a}\bar{b}}$ are smooth functions of $(r^2, X^{\bar{a}})$.
Then exactly as for the Euclidean continuation of the static problem, we may make the periodic identification $\bar{\tau} \sim \bar{\tau} + 2 \pi \kappa$ so that this auxiliary geometry $g^{(aux)}$ is smooth without boundary at $r=0$ -- the apparent boundary associated to the horizon points $r = 0$ which form $\partial \bar{\Omega}_{\mathcal{H}}$ is only a coordinate singularity in this auxiliary Riemannian space where the circle generated by $\partial/\partial \bar{\tau}$ degenerates, and may be removed in the usual way by moving to new `Cartesian' coordinates, the Euclidean version of \eqref{eq:horizcoord}; 
\begin{equation}
\label{eq:horizcoord2}
\sigma = r \cos{ \kappa \, \bar{\tau}} \; , \quad \rho = r \sin{ \kappa \, \bar{\tau}} \; .
\end{equation}
With such a periodic $\bar{\tau}$, regularity of a scalar function (not necessarily symmetric under $\partial/ \partial \bar{\tau}$) on this auxiliary space requires it is a smooth function of $(r^2, \bar{\tau}, X^{\bar{a}})$, and regularity of a covector field $u_\mu$ implies it behaves as,
\begin{equation}
u = r^2 \,u_{\bar{\tau}} \,d\bar{\tau} + r\, u_r \,dr + u_{\bar{a}} \,dX^{\bar{a}}
\end{equation}
with component functions $u_{\bar{\tau}}$, $u_r$ and $u_{\bar{a}}$ being smooth functions of $(r^2,\bar{\tau}, X^{\bar{a}})$.

Now consider the following elliptic equation on the auxiliary space,
\begin{align}
\label{eq:aux2}
\left( {\nabla}^2_{(aux)}  +   {u}^\mu   \partial_\mu \right)  F  & \ge - 2 \, \Lambda \, F  
\end{align}
for $F$ a smooth function and $u$ a smooth covector field, with the expression being written covariantly in the auxiliary metric $g^{(aux)}$. For $\Lambda \le 0$ the maximum principle applied to this equation implies that $F$ is either constant, or if not, cannot attain a maximum at any interior point of the auxiliary geometry. The important point is that, with suitably periodic $\bar{\tau}$, points corresponding to the original spacetime points in the Killing horizon are now interior points, and so the maximum principle applies at these. 
Now let us consider the special case that $F$ is independent of the Killing directions, $\frac{\partial}{\partial y^A} F = 0$. Hence it reduces to a scalar on $\bar{\Omega}$.
Let us also take the smooth covector field $u$ to be,
 \begin{align}
u = \bar{u}_a \,dX^a \; , \quad \bar{u}_a = \bar{v}_a +\frac{1}{2} \partial_a \log\left( 1 + \frac{1}{N} \,\bar{g}^{ab} A_a A_b  \right) \, ,
\end{align}
where $\bar{v}$ is the same smooth covector as in~\eqref{eq:ineqstationary2}.
Noting from~\eqref{eq:spacetimenormal} that in normal coordinates at the horizon we have $N \sim O(r^2)$ and $A_a \sim O(r^2)$, with $\bar{g}^{ab} \sim O(r^0)$ then indeed $u$ is smooth at the base points $\partial \bar{\Omega}_{\mathcal{H}}$. Then the above equation~\eqref{eq:aux2} can be written covariantly over $\bar{\Omega}$, where,
\begin{align}
\label{eq:auxbase}
\left( \bar{\nabla}^2 + \bar{J}^a_{(aux)} \partial_a + \bar{u}^a  \partial_a + \bar{v}^a  \partial_a \right) F  & \ge - 2 \, \Lambda \, F \; , \quad \bar{J}^{(aux)}_a = \frac{1}{2} \, \partial_a \log N \; ,
\end{align}
where $\bar{\nabla}$ is again the covariant derivative of $\bar{g}$. While $\bar{u}$ and $\bar{v}$ are smooth, including at the points $\partial \bar{\Omega}_{\mathcal{H}}$,  $\bar{J}^{(aux)}$ is not. Recall we know that $F$ cannot attain a non-negative maximum at the points $\partial \bar{\Omega}_{\mathcal{H}}$ that correspond to horizons in the original spacetime. However, noting that,
\begin{align}
\bar{J}_a = \bar{J}^{(aux)}_a + \bar{u}_a 
\end{align}
then we also see that the above equation~\eqref{eq:auxbase} is precisely the same as~\eqref{eq:ineqstationary2} under the replacement $F \to \omega$. Furthermore since the smoothness constraint on our choice of $F$ is precisely the same as the smoothness constraint on $\omega$ (i.e., smooth and even in the normal coordinate $r$ of~\eqref{eq:auxnormal}) then since $F$ cannot attain a non-negative maximum at $r = 0$ we see neither can $\omega$ attain a non-negative maximum at a point in a horizon, which is what we desired to show.

Thus in summary, considering~\eqref{eq:ineqstationary}, the function $\omega$ is either constant, or if not, cannot attain a non-negative maximum value in the interior of the base $\mathcal{M}$, or as we have now learned, at boundaries of $\partial \mathcal{M}$ which correspond to spacetime axes of symmetry and horizons. Following our previous discussion this leads to the conclusion that given our asymptotics then $\omega \le 0$, and then the statements in our {\bf Claim} then follow.

%
\subsubsection*{Kerr example}
\label{subsec:Kerr}
%

In order to make the discussion above more concrete, we feel it is useful to have the example of the Kerr solution to illustrate the decomposition of the metric over a regular base $\mathcal{M}$, and also see the explicit construction of the auxiliary space with metric $g^{(aux)}$. Recall the Kerr metric in Boyer-Lindquist coordinates is already of ``$t$-$\phi$'' symmetric form, with fibre metric,
\begin{align}
\label{eq:Kerrfibre}
G_{tt}  = - \frac{ \Delta - a^2 \sin^2{\theta} }{ \Sigma} \; , \quad
G_{\phi\phi}  = \sin^2{\theta} \frac{ ( r^2 + a^2 )^2 - \Delta a^2 \sin^2{\theta} }{ \Sigma} \; , \quad
G_{t\phi} = G_{\phi t} = - a \sin^2{\theta} \frac{ r^2 + a^2 - \Delta }{\Sigma} 
\end{align}
and base,
\begin{align}
\label{eq:Kerrbase}
\hat{h}_{ij} dx^i dx^j & = \Sigma \left( d\rho^2 + d\theta^2 \right)
\end{align}
where $\Delta = r^2 + a^2 - 2 M r$, $\Sigma = r^2 + a^2 \cos^2{\theta}$ and we have defined $r = M + \sqrt{M^2 -a^2} \cosh{\rho}$, so that $\rho = 0$ corresponds to the outer Kerr horizon $r = r_h$ where $\Delta$ vanishes. Note that in the coordinates $\rho, \theta$ the base is regular with boundaries at $\rho = 0$ (the horizon) and $\theta = 0, \pi$ (the axis of symmetry).

To move to the auxiliary space over a small enough neighbourhood $\Omega_p$ of a point $p$ \emph{outside} the ergoregion (so $G_{tt} < 0$) we write the full spacetime in the form \eqref{eq:spacetime},
\begin{align}
\label{eq:Kerrfull1}
ds^2 =  G_{tt} \, dt + 2 G_{t\phi} \, dt \, d\phi  + \Sigma \left( d\rho^2 + d\theta^2 \right) + G_{\phi\phi} \, d\phi^2 
\end{align}
so then,
\begin{align}
\label{eq:Kerraux1}
ds^2_{(aux)} =  \left| G_{tt} \right| d\bar{\tau}^2 + \Sigma \left( d\rho^2 + d\theta^2 \right) + G_{\phi\phi}  \, d\phi^2 
\end{align}
and indeed the constant $\bar{\tau}$ section of this is regular and smooth at axes of symmetry. 

For a point $p$ in the neighbourhood of the horizon we must first move to a co-rotating frame, transforming the fibre coordinates as, $\tilde{t} = t$ and $\tilde{\phi} = \phi - \Omega t$, where  $\Omega$ is the angular velocity of the horizon, so $\partial / \partial \tilde{t}$ is its null generator. Then $a$ and the surface gravity $\kappa$ (with respect to the null generator $\partial / \partial \tilde{t}$) are determined from $M$ and $\Omega$ as,
\begin{align}
a = \frac{4 M^2 \Omega}{1 + 4 M^2 \Omega^2} \; , \quad \kappa = \frac{1 - 4 M^2 \Omega^2 }{4 M} \; .
\end{align}
Now $G_{\tilde{t}\tilde{t}}<0$ in the neighbourhood of the horizon, and $G_{\tilde{t}\tilde{t}}=0$ on the horizon. Then we may take an auxiliary Riemannian space, 
\begin{align}
\label{eq:Kerraux2}
ds^2_{(aux)} =  \left| G_{\tilde{t} \tilde{t} } \right| d\bar{\tau}^2 + \Sigma \left( d\rho^2 + d\theta^2 \right) + G_{\tilde{\phi} \tilde{\phi} }\, d\tilde{\phi}^2 
\end{align}
and making the identification $\bar{\tau} = \bar{\tau} + 2  \pi \kappa$ this is indeed regular with no boundary at $\rho = 0$. Indeed one can check near the horizon we have,
\begin{align}
ds^2_{(aux)} = \left( \frac{4 M \left( M + 2 M^3 \Omega^2 + 2 M^3 \Omega^2 \cos{2 \theta} \right)}{\left(1 + 4 M^2 \Omega^2 \right)^2} \left( dX'^2 + dY'^2 + d\theta^2 \right) + \frac{4 M^2 \sin{\theta}^2}{1 + 2 M^2 \Omega^2 + 2 M^2 \Omega^2 \cos{2 \theta}}  d\tilde{\phi}^2 \right) \left(1 + O\left( \rho^2 \right) \right)
\end{align}
where $X' = \rho \cos \kappa \bar{\tau}$, $Y' = \rho \sin \kappa \bar{\tau}$. For a point $p$ near the ergosurface we may find a suitable linear combination of $\partial/\partial t$ and $\partial / \partial \tilde{t}$ that is a timelike Killing vector to construct the auxiliary geometry.

%
\section{Application to finding numerical black hole solutions}
\label{sec:application}
%

We now consider the implications of this result to the original motivating problem of finding black holes numerically using the harmonic approach detailed in section \S\ref{sec:reviewmethod}. If we only have cosmological constant matter and wish to find ``$t$-$\phi$'' symmetric spacetimes, constructing a reference connection as described, then in the end we must solve the Ricci soliton equation \eqref{eq:harmonic} wishing to find Einstein solutions rather than solitons. Following equation \eqref{eq:xiorthog}, the vector $\xi$ is orthogonal to the Killing vectors associated to the ``$t$-$\phi$'' symmetry, and hence the results of our {\bf Claim} apply taking $v$ to be $\xi$. Thus we may guarantee no soliton solutions exist in the cases $\Lambda \le 0$ provided the asymptotics of the solution and vector $\xi$ obey the conditions surrounding equation \eqref{eq:defn}, and the soliton charge $Q$ vanishes. 

Let us consider separately the two situations, $\Lambda = 0$ and $\Lambda < 0$, and assume we are interested in solutions which are asymptotically flat in the first case, and asymptotically global-AdS in the second. We note that more general asymptotics are possible -- for example, asymptotically Kaluza-Klein in the first case, or asymptotically \emph{locally} AdS in the second. However our considerations in these basic settings will suffice to understand more general situations of interest. The question is then whether the asymptotic requirements to rule out the existence of unwanted soliton solutions are automatically satisfied by these asymptotic behaviours when the reference connection is constructed taking the same behaviour, as outlined in \S\ref{sec:reviewmethod}, or whether some additional condition must be imposed, or whether perhaps there is an inconsistency and such asymptotic requirements on $\xi$ cannot be imposed -- and hence the formal result of section~\S\ref{sec:nonexistence} is of no relevance to the numerical setting we are motivated by. Happily we shall find that indeed either the condition is automatic, or may be imposed as a reasonable boundary condition.

%
\subsubsection{$\Lambda = 0$ and the asymptotically flat case}
\label{subsec:flat}
%

A spacetime with dimension $D\ge 4$ is asymptotically flat if for  sufficiently large $R$ the metric behaves as,
\begin{align}
\label{eq:asymflat}
ds^2 & = -dt^2 + \delta_{ij} dx^i dx^j + a_{\mu\nu}(x^\alpha)  \;, \quad  \forall \; \delta_{ij} x^i x^j > R^2 \,,\nonumber \\
a_{\mu\nu} & \sim O\left( \frac{1}{R^{D-3}} \right) \,, \nonumber \\
\partial_\alpha a_{\mu\nu} & \sim O\left( \frac{1}{R^{D-2}} \right) \,, \nonumber \\
\partial_\alpha \partial_\beta a_{\mu\nu} & \sim O\left( \frac{1}{R^{D-1}} \right) \, .
\end{align}
Provided we choose the same behaviour for the reference metric one finds, $\xi_\mu = O\left( \frac{1}{R^{D-2}} \right) $
and indeed the scalars $\phi = O\left( \frac{1}{R^{2(D-2)}} \right)$ and $\omega = O\left(\frac{1}{R^D} \right)$  vanish asymptotically, and sufficiently fast, as we required in the setting where our {\bf Claim} applies. However, the $O(1/R^{D-2})$ term in the asymptotic expansion of $\xi_\mu$ gives rise to a finite soliton charge $Q$.

Since we are only interested in solutions to the harmonic Einstein's equation, we might wonder whether for these the finite scalar charge is forced to vanish by the equations of motion. However this is not the case as we may illustrate with a simple counterexample.
Consider a general static and spherically symmetric metric in four dimensions (higher dimensions are analogous),
\begin{equation}
ds^2=-T(r)\,dt^2 + A(r)\,dr^2 + r^2\,S(r)\,d\Omega_{(2)}^2\,.
\label{eq:AFexample}
\end{equation}
Now impose that \eqref{eq:AFexample} is asymptotically flat as above, so,
\begin{equation}
T(r) = 1+\frac{t_1}{r}+O\left(\frac{1}{r^2}\right)\,,\quad A(r) = 1 + \frac{a_1}{r} + O\left(\frac{1}{r^2}\right)\,,\quad S(r) = 1 + \frac{s_1}{r} + O\left(\frac{1}{r^2}\right)\,,
\label{eq:AFexpansion}
\end{equation}
where $t_1$, $a_1$ and $s_1$ are constants. To proceed, we choose a reference metric which is also static and spherically symmetric, and asymptotically flat in the above sense. We shall denote the corresponding coefficients in the near infinity expansion with a bar $\bar\,$. Then the leading order term in the non-vanishing component of the one-form $\xi_\mu$ is given by
\begin{equation}
\xi_r = \frac{1}{2\,r^2}\left(3\,a_1 - 3\,\bar a_1-2\,s_1+2\,\bar s_1 + t_1 - \bar t_1\right) + O\left(\frac{1}{r^3}\right)\,.
\end{equation}
giving a finite scalar charge $Q = \lim_{r \to \infty} 4 \pi r^2 \xi_r$.

Now assume our spacetime solves the harmonic Einstein's equation \eqref{eq:harmonic}. Then the coefficients in the $1/r$ expansion are constrained, and to leading order one finds,
\begin{equation}
s_1 = \bar s_1 + a_1 - \frac{1}{2}\left(3\,\bar a_1 + \bar t_1\right) \,,
\label{eq:conda1}
\end{equation}
which implies the scalar charge for a solution of the harmonic Einstein's equation is,
\begin{equation}
Q = 2 \pi \left( t_1 + a_1 \right) \,.
\end{equation}
From the Newtonian expansion of general relativity we famously know that static spherically symmetric asymptotically flat solutions have $t_1 = - a_1$. Indeed we see that our scalar charge is in fact measuring the deviation of the PPN parameter `$\gamma$' of the spacetime from its general relativity value of one.

Thus we see that asymptotically flat boundary conditions for $\Lambda = 0$ are sufficient to guarantee $\phi, \omega \to 0$, and that the scalar charge $Q$ is finite, but not that it automatically vanishes. However in solving the numerical system we could impose the further condition on our asymptotic behaviour that $Q$ vanishes -- for example, in the simple spherically symmetric context, impose that $t_1 = - a_1$, and solve the harmonic equations only in that class of spacetimes. Then if a solution is found, our results imply it cannot be a soliton. 

It is interesting to contrast this with the static case where simply having $\phi \to 0$ was sufficient to rule out solitons. Here we see in the stationary case that we must impose the asymptotics are not only flat, but also have vanishing soliton charge. This is an entirely reasonable, indeed sensible thing to do, but does hint that perhaps a stronger result than we have obtained may still be possible in the $\Lambda = 0$ case.

%
\subsubsection{$\Lambda < 0$ and the asymptotically global-AdS case}
\label{subsec:AdS}
%

Let us now consider $D$-dimensional ``$t$-$\phi$" reflection symmetric solutions of the Ricci soliton equation~\eqref{eq:harmonic} that are asymptotic to global AdS, and have a negative cosmological constant, $\Lambda = - \frac{D-1}{\ell^2}$. Here $\ell$ is the AdS length and for convenience we will choose units so that $\ell =1$. We restrict attention to the cases  $D\ge4$. We will now show that for appropriate choices of reference metrics, the soliton charge $Q$ must vanish on any solution to equation~\eqref{eq:harmonic}, and thus such a solution must be an Einstein metric.

Whilst we may present such a solution in Fefferman-Graham gauge, we must be careful in the analysis that follows that we may only choose such coordinates for the metric or for the reference metric, but not both simultaneously. In Fefferman-Graham gauge, $D$-dimensional global AdS can be written as (see eg. \cite{Marolf:2013ioa}),
\begin{equation}
ds^2 = \frac{1}{z^2}\left[ dz^2 - \left( 1+\frac{1}{2}\,z^2 +\frac{1}{16}\,z^4 \right) dt^2 +  \left( 1-\frac{1}{2}\,z^2 +\frac{1}{16}\,z^4 \right) \sigma_{ab}\,dx^a\,dx^b \right]\,,
\label{eq:globalAdSFG}
\end{equation}
where $\sigma_{ab}$ is the metric on the round unit $(D-2)$-sphere. 
We write the solution to \eqref{eq:harmonic} in the form,
\begin{equation}
ds^2 = \frac{1}{z^2}\left[ N(dz + A_\alpha\,dx^\alpha)^2 + h_{\alpha\beta}\,dx^\alpha\,dx^\beta \right]\,,
\label{eq:AdSmetric}
\end{equation}
where the boundary of AdS is located at $z=0$ and we allow the metric coefficients $N$, $A_\alpha$ and $h_{\alpha\beta}$  to depend on the spatial coordinates. In our ``$t$-$\phi$'' reflection symmetric context they should be invariant in the Killing vector directions $\Psi$, although we will not need to use that explicitly here.

We write down the (``$t$-$\phi$'' reflection symmetric) line element of the reference metric in the same form, with the replacement $N \to \bar{N}$, $A_\alpha \to \bar{A}_\alpha$ and $h_{\alpha\beta} \to \bar{h}_{\alpha\beta}$.
We will chose a reference metric that is asymptotic to, although not necessarily equal to, global AdS. We adapt our coordinates to this reference metric, choosing them so that it is in Fefferman-Graham gauge, $\bar N  = 1$ and  $\bar A_\alpha  = 0$ , and we take,
\begin{equation}
\begin{aligned}
&\bar h_{tt} = 1+\frac{1}{2}\,z^2 +\frac{1}{16}\,z^4+ z^d\,\bar h^{(d)}_{tt}(x) + O(z^{d+1})\,,\\
&\bar h_{ta} =  z^d\,\bar h^{(d)}_{ta}(x) + O(z^{d+1})\,,\\
&\bar h_{ab} = \sigma_{ab}\left(1-\frac{1}{2}\,z^2 +\frac{1}{16}\,z^4+ z^d\,\bar h^{(d)}_{ab}(x) + O(z^{d+1})\right)\,,\\
\end{aligned}
\end{equation}
where $d = D-1$ is the number of boundary spacetime dimensions. 
For the spacetime metric we begin with a general expansion in $z$, 
\begin{equation}
\begin{aligned}
&N = 1 + \sum_{i=1}^\infty z^i\,n^{(i)}(x) \,,\\
&A_\alpha = \sum_{i=1}^\infty z^i\, a^{(i)}_\alpha(x) \,,\\
&h_{tt} = 1+\frac{1}{2}\,z^2 +\frac{1}{16}\,z^4 + \sum_{i=1}^\infty z^i\, h^{(i)}_{tt}(x)\,,\\
&h_{ta} = \sum_{i=1}^\infty z^i\, h^{(i)}_{ta}(x)\,,\\
&h_{ab} = \sigma_{ab}\left(1 -\frac{1}{2}\,z^2 +\frac{1}{16}\,z^4+ \sum_{i=1}^\infty z^i\, h^{(i)}_{ab}(x)\right)\,,
\end{aligned}
\label{eq:expansionAdS}
\end{equation} 
and then solving \eqref{eq:harmonic} order by order in $z$ one finds the terms, $n^{(i)}$, $a^{(i)}_\alpha$, $h_{\alpha\beta}^{(i)}$ all vanish for $i < d$. 
The functions $n^{(d)}$, $a^{(d)}_\alpha$ and $h^{(d)}_{\alpha\beta}$ determining the leading behaviour of the expansion are the free data specifying the solution, but are constrained by equation \eqref{eq:harmonic} to obey,
\begin{equation}
\begin{aligned}
&n^{(d)} = -  h^{(d) \alpha}_\alpha = \frac{d-2}{2} \bar h^{(d) \alpha}_\alpha \,,\\
&a^{(d)}_\alpha = 0\,, \\
& D^\alpha h^{(d)}_{\alpha\beta} = \frac{d+2}{2} D^\alpha \bar{h}^{(d)}_{\alpha\beta} - \partial_\beta \bar{h}^{(d) \alpha}_{\alpha}  \, .
\end{aligned}
\end{equation}
Here indices are raised and lowered with respect to the boundary spacetime $-dt^2 + \sigma_{ab}\,dx^a\,dx^b$, and $D^\alpha$ is its covariant derivative. 
Then the integrand for the soliton charge defined in equation~\eqref{eq:AdScharge} is,
\begin{equation}
\begin{aligned}
&dS^\mu \xi_\mu = \frac{1}{2} \sqrt{\sigma} \left( d \, n^{(d)} + (d-2)  \textrm{tr}\left( h^{(d)} - \bar h^{(d)} \right) \right) + O(z) = O(z) \, .
\end{aligned}
\label{eq:AdScharge}
\end{equation}
and hence vanishes on a solution to~\eqref{eq:harmonic} which therefore has  $Q = 0$. Furthermore one finds $\phi \sim O(z^{2 d+2})$ and $\omega \sim O(z^{d+1})$ so these scalars have sufficiently quick fall off to satisfy the conditions in our {\bf Claim}.
As a consequence, without imposing any further asymptotic conditions we deduce that no  ``$t$-$\phi$'' reflection symmetric soliton solutions can exist.

Thus in the asymptotically global-AdS case with reference metric chosen as above, $Q = 0$ automatically for any ``$t$-$\phi$'' reflection symmetric solution of the harmonic  Einstein's equation, and this does not need to be imposed as an additional asymptotic condition. So for the same boundary conditions as in the static case, we find no Ricci soliton solutions can exist to the harmonic Einstein's equation.

%
\section{Summary}
\label{sec:discussion}
%

In this paper we have provided a new argument for the non-existence of Ricci soliton solutions to the harmonic Einstein's equation \eqref{eq:harmonic} with cosmological constant matter $\Lambda \le 0$  for stationary spacetimes with appropriate asymptotic behaviour and ``$t$-$\phi$'' reflection symmetry.
We have discussed how the previous static non-existence result of \cite{Lucietti} cannot be simply extended to the stationary case, and our new stationary argument is somewhat more subtle. In particular we show for a ``$t$-$\phi$'' reflection symmetric solution to the soliton equation we define a quantity $Q$, the soliton charge of the solution, and show that a non-trivial steady ($\Lambda = 0$) or expanding ($\Lambda < 0$) soliton has $Q < 0$, and if $Q = 0$ the solution must be Einstein. 
We then apply this to solutions of the harmonic Einstein's equation. 
Taking $\Lambda = 0$ and requiring asymptotically flat solutions we find that for our choice of reference connection one must impose $Q = 0$ to rule out soliton solutions. Since all Einstein solutions must have $Q = 0$ this is a reasonable asymptotic condition to impose.
In the case $\Lambda < 0$ and requiring asymptotically global-AdS solutions, we find that the usual choice of reference metric is sufficient to ensure that $Q = 0$ and no soliton solutions to the harmonic Einstein's equation can exist.
 
Thus motivated by giving explicit methods to find solutions to the Einstein's equation in physically exotic settings we have found a nice result. At least for zero or negative cosmological constant matter, and appropriate choices of reference connections and asymptotic conditions, the only solutions to the harmonic Einstein's equation are the ones we wish to find. One must analyse the asymptotics to see whether $Q = 0$ automatically or not, but perhaps it is always sensible to impose $Q = 0$ as an asymptotic condition given the solutions desired must have that property.
 
As we have noted above, in the  static non-existence result of \cite{Lucietti} for $\Lambda = 0$ there was no requirement to have vanishing soliton charge to ensure non-existence of solitons -- asymptotic flatness of the metric and reference metric were sufficient. However in our new stationary result we must additionally impose vanishing soliton charge to ensure no solitons. We feel it is likely then that our result can be improved so that this condition on soliton charge is not required in the stationary case. It would be interesting to explore this further. Another interesting direction is to examine the existence of solitons in the presence of matter fields -- to date even in the static case this has not yet been addressed.
Similarly, it would be desirable to extend our arguments to other spacetimes containing extremal horizons or non-trivial boundaries such as Randall-Sundrum branes \cite{Randall:1999ee}.

%
\section*{Acknowledgements}
%

We are grateful for the hospitality provided by the BIRS ``Geometric flows: Recent developments and applications" program where this work was initiated. TW's work was supported by the STFC grant ST/J0003533/1. PF is supported by a Royal Society University Research Fellowship and by the H2020 ERC Starting Grant ``New frontiers in numerical relativity", grant agreement No. NewNGR-639022.

%
\bibliographystyle{apsrev4-1}
\bibliography{paper}
%

\end{document}